\documentclass[twocolumn,letterpaper]{IEEEAerospaceCLS}  

\usepackage[]{graphicx}    
\usepackage{braket}
\usepackage{amsmath}
\usepackage{amsfonts}
\usepackage{enumitem}
\usepackage{mathtools}
\usepackage{blkarray, bigstrut} 
\usepackage{bbm}   
\usepackage{paralist}
\usepackage{subfigure}
\usepackage{caption}
\usepackage{siunitx}

\newcommand{\ignore}[1]{}  

\newcommand{\X}{{\mathcal{X}}}      
\newcommand{\Y}{{\mathcal{Y}}}      

\newcommand{\costQ}{{\mathbf{\Gamma}}}
\newcommand{\kalmanQ}{{Q_{{\text{proc}}}}}
\newcommand{\given}{\,|\,}
\newcommand{\givennn}{|} 
\newcommand{\kalmanPzero}{P_{0 \givennn 0}}

\newcommand{\fov}{|\text{FoV}|}

\newcommand{\meas}{y}

\newcommand{\xkn}{{{{x}}^n_k}}
\newcommand{\yk}{{{\meas}_k}}
\newcommand{\ykone}{{{{\meas}}^1_k}}
\newcommand{\ykMk}{{{{\meas}}^{M_k}_k}}

\DeclareFontFamily{OT1}{pzc}{}
\DeclareFontShape{OT1}{pzc}{m}{it}{<-> s * [0.900] pzcmi7t}{}
\DeclareMathAlphabet{\mathpzc}{OT1}{pzc}{m}{it}
\newcommand{\scx}{{\mathpzc{x}}}
\newcommand{\scy}{{\mathpzc{y}}}

\newcommand{\WrFromKrooks}{W_r}   
\newcommand{\WcFromKrooks}{W_c}   
\newcommand{\thetaRKrooks}{\theta_r}
\newcommand{\thetaCKrooks}{\theta_c}
\newcommand{\WrFromMTDA}{W'_r}   
\newcommand{\WcFromMTDA}{W'_c}   
\newcommand{\thetaRMTDA}{\theta'_r}
\newcommand{\thetaCMTDA}{\theta'_c}
\newcommand{\texp}{{\mathbb{T}}\,\textrm{exp}}

\newcommand{\qkrooks}{{q_{\scriptscriptstyle \rm{kR}}}}
\newcommand{\qqkrooks}{{Q_{\scriptscriptstyle \rm{kR}}}}
\newcommand{\qmtda}{{q_{\scriptscriptstyle \rm{MTDA}}}}
\newcommand{\qqmtda}{{Q_{\scriptscriptstyle \rm{MTDA}}}}

\begin{document}
\title{Multiple Target Tracking and Filtering using Bayesian Diabatic Quantum Annealing}

\author{%
Timothy M. McCormick, Zipporah Klain, Ian Herbert,\\ Anthony M. Charles, R. Blair Angle, Bryan R. Osborn, Roy L. Streit \\ 
Metron, Inc.\\
1818 Library St., Suite 600\\
Reston, VA 20190\\
\{McCormickT, Klain, Herbert, CharlesA, Angle, Osborn, Streit\}@metsci.com
\thanks{\footnotesize 978-1-6654-3760-8/22/$\$31.00$ \copyright2022 IEEE}              
}

\maketitle

\thispagestyle{plain}
\pagestyle{plain}

\begin{abstract}
In this paper, we present a hybrid quantum/classical algorithm to solve an NP-hard combinatorial problem called the multiple target data association (MTDA) and tracking problem. We use diabatic quantum annealing (DQA) to enumerate the low energy, or high probability, feasible assignments, and we use a classical computer to find  the Bayesian expected mean track estimate by summing over these assignments.  We demonstrate our hybrid quantum/classical approach on a simple example. This may be the first demonstration of a Bayesian hybrid quantum-classical
multiple target tracking filter.   We contrast our DQA method with the adiabatic quantum computing (AQC) approach to MTDA. We  give a  theoretical overview of DQA and characterize some of the technical limitations of using quantum annealers in this novel diabatic modality. 
\end{abstract}

\tableofcontents
\section{Introduction}

To compute  the (Bayesian) posterior distribution in  multi-target  tracking and data association (MTDA) problems, we need to evaluate a probabilistic sum, each term of which is conditioned on exactly one  feasible assignment of measurements to targets.  Enumerating the feasible assignments is an NP-hard combinatorial problem.

The fundamental quantity of interest in Bayesian inference  is the posterior distribution. Point estimates, when needed,  are ``extracted'' from the posterior by Bayesian decision theoretic methods (e.g., minimum risk). Two  common point estimators are the  MAP (maximum a posteriori) and the mean of the posterior distribution, both of which are used in tracking. 

Adiabatic quantum computing (AQC) methods are used in \cite{mtda}, \cite{aeroMtda}, \cite{zaech} to find the optimal assignment. Conditioned on this assignment, target  tracks are computed by classical methods. In contrast, Bayesian methods do not select the best assignment, but instead sum over  all the assignments. Thus, Bayesian methods are better  suited to real world problems that  often have high false alarm rates and low target detection probabilities.    

This paper shows that diabatic quantum annealers (DQA) can be used to compute the Bayesian mean target state estimator.  
DQA finds a collection of low-energy configurations that are  near but not necessarily in the ground state. Thus, in sharp contrast to AQC, DQA deliberately violates the adiabatic  condition in order to anneal to a set of low-energy configurations.  The DQA-generated collection of  low-energy states is then passed to a classical computer to compute the mean state estimator.

Finding the low-energy configurations is a high computational complexity problem on a classical computer, but computing the mean state given the low-energy configurations is not.  This is consistent with the over-arching strategy of exploiting QC  methods to solve the NP-hard part(s)  of a problem, which are often combinatorial, and then coupling the QC output with a classical computer to calculate the  target tracks.    

Section \ref{dataassoc} deals with the mathematical details of the JPDA tracking filter.  Section \ref{isingSec}  discusses the transformation of the MTDA problem into an Ising model suitable for calculating on a QC such as D-Wave's 2000Q.  Section \ref{annealSec} discusses  using DQA to find low-energy non-ground states.  Section \ref{resultsSingle} presents results.  Section \ref{resultsRecur} gives results of using DQA to find  feasible configurations, i.e., assignments. To our knowledge, this is the first demonstration of a Bayesian hybrid quantum-classical multiple target tracking filter. 
Section \ref{ConcludingRemarks} summarizes our findings about the utility of DQA for Bayesian inference.

\section{Data Association and Tracking}\label{dataassoc}
The joint probabilistic data association (JPDA) filter \cite{ybsANDli}  is a classical Bayesian 
target filter that estimates the joint posterior PDF for a known number of targets, $N$, by
fusing multiple target-measurement assignment hypotheses in a principled manner.
The JPDA filter is NP-hard because the number of assignments grows rapidly with problem size. 

In the JPDA model, targets are assumed to be independent of one another and causally independent
of the measurement process. At each scan $k$, a sensor produces a set of $M_k \geq 0$ point 
measurements, each of which is either
target-induced or the result of an independent clutter (false alarm) process.
A given target generates at most one sensor measurement per scan, 
and any given measurement is assigned to exactly one of the targets or to the clutter process.
All point measurements are superposed in a common measurement space $\Y$.

The JPDA model is inherently flexible and is amenable to particle filter (sequential Monte Carlo) methods.
As the primary focus of this paper is target-measurement assignments, the JPDA model employed
will adopt all of the original/standard JPDA assumptions. These simplifying assumptions, 
described below for a specific scenario, allow for an analytical solution of the posterior PDF, 
namely, a multivariate Gaussian.

In the tracking example  in this paper,  units of length are in meters, and time units are in seconds. 
Scans occur at one second intervals, beginning at time $t_1 = 1$; i.e., $\Delta t = 1$. 
The $k$-th scan occurs at time $t_k = k$ 
(the scan time is defined as the end time of the scan interval). 
There are $N=4$ targets, with identical state spaces:
$\X^n \equiv \X \subset {\mathbb{R}}^{4}$,   $n=1,\ldots,N$.
The  state vector for target $n$ at scan $k$ comprises two spatial components,
$\scx$ and $\scy$, and two velocity components, $\dot{\scx}$ and $\dot{\scy}$,
and is denoted $\xkn = \left(\scx^n_k, {\dot{\scx}}^n_k, \scy^n_k, {\dot{\scy}}^n_k\right)$.

At the reference time, $t_0 = 0$, no measurements are available and
target $n$ is assumed to have prior PDF $\mu_0^n(x_0^n)  = {\mathcal{N}}\! \big(x_0^n ; \hat{x}_{{0 \givennn 0}}^n , \kalmanPzero^n \big)$, where
 ${\mathcal{N}}(x ; \mu , \Sigma)$ represents the PDF of a multivariate Gaussian with mean vector $\mu$ and covariance matrix $\Sigma$ evaluated at $x$. 
Due to the assumption of target independence, the joint prior PDF for all target states is the product over $n$ of the marginals $\mu_0(x_0^n)$, that is,
\begin{equation}    \label{eq:jpdaPrior}  
p_0(x_0^1,\ldots,x_0^N) = \prod_{n=1}^N \mu_0^n(x_0^n) = \prod_{n=1}^N {\mathcal{N}}\!\left(x_0^n ; \hat{x}^n_{{{0 \givennn 0}}} ,\,
   P_{0 \givennn 0}^n \right).
\end{equation}
As will be seen, the JPDA filter imposes this factored form at each step $k$ of the  recursion.  This is an approximation, and it  ensures the posterior distribution has the exact same mathematical form as the prior distribution.  In other words, the approximation  closes the Bayesian  recursion.

Targets move according to the linear-Gaussian motion model 
\begin{equation}     \label{eq:jpdaMotUpdate}    
    p(x_k^n \given x_{k-1}^n)   =  {\mathcal{N}}\!\left(x_k^n ; F x_{k-1}^n , \kalmanQ \right) ,
\end{equation}
where the process (motion) matrix $F$ and the process noise covariance matrix $\kalmanQ$
are given by 
$$     
F = 
\begingroup 
\setlength\arraycolsep{2.6pt}
\begin{pmatrix}
1 \,  & \Delta t   & 0   & 0 \\
 0  \,   &  1   & 0   & 0 \\
0 \,  & 0   & 1   & \Delta t  \\
 0  \,   &  0   & 0   & 1
\end{pmatrix}  
\endgroup
, \,
\kalmanQ  = 
\begingroup 
\setlength\arraycolsep{2.6pt}
\sigma^2_p\begin{pmatrix}
 \frac{\Delta t^3}{3} &  \frac{\Delta t^2}{2} & 0 & 0 \\
  \frac{\Delta t^2}{2} &  \Delta t  & 0 & 0 \\
  0 & 0 & \frac{\Delta t^3}{3} &  \frac{\Delta t^2}{2} \\
  0 & 0 & \frac{\Delta t^2}{2} &  \Delta t  
\end{pmatrix} .
\endgroup
$$
We take $\sigma_p = 3$.  These matrices are independent of scan index $k$.

At each scan, target $n$ generates a sensor measurement with (constant)
probability of detection $p_d = 0.9$.
The (bounded) measurement space $\Y = [-600 \textrm{m},600 \textrm{m}] \times [-600 \textrm{m},600 \textrm{m}] \subset {\mathbb{R}}^{2}$
comprises the two spatial components $\scx$ and $\scy$.
Given a measurement $y = (\scx,\scy)$ induced by target $n$ at scan $k$,
the measurement likelihood has the following linear-Gaussian form:
\begin{equation} \label{eq:jpdaLF}
p(y \given x_k^n) = {\mathcal{N}}\!\left(y ; H x_k^n , R \right) , 
\end{equation}
where the measurement matrix $H = 
\begingroup 
\setlength\arraycolsep{2.6pt}
\begin{pmatrix}
 1 & 0 & 0 & 0 \\
 0 & 0 & 1 & 0
\end{pmatrix}
\endgroup$
extracts the two spatial components of $\xkn$,
and $R = \sigma^2_M
\begingroup 
\setlength\arraycolsep{2.6pt}
\begin{pmatrix}
 1 & 0 \\
 0 & 1
\end{pmatrix}
\endgroup$,
$\sigma_M = 25\ \textrm{m}^2$, is the measurement covariance.

The clutter model employed is a homogeneous Poisson point process (PPP)
with mean $\lambda = 1$. That is, at each scan the number, $n_c$, of clutter points 
is generated according to a Poisson probability mass function (PMF) with mean $\lambda$.
If $n_c>0$, the clutter points are then uniformly and independently distributed over the 
the entire measurement space (field of view) $\Y$, the area of which is denoted 
$\fov = 1.44\times 10^6\ \textrm{m}^2$.

Let $\yk = \{\ykone,\ldots,\ykMk\}$ be the measurement set at scan $k$.
If the target-measurement assignment is known, then the exact Bayesian posterior
distribution is obtained via $N$ independent Kalman filter updates (see \cite{ybsANDli}) and
is of the same form as the prior given in Eqn. (\ref{eq:jpdaPrior}), namely,
the product of $N$ independent Gaussians.
However, since the assignments are unknown, the exact Bayesian posterior
 is a weighted sum over all feasible assignment matrices $S$
(see Eqn. (7) in Ref. \cite{aeroMtda}), 
i.e., a (large)
Gaussian mixture of the form
\begin{equation}   \label{eq:jpdaExactPost}
p_k(x_k^1,\ldots,x_k^N | y_k)    =    \sum_{S}   
  \text{Pr}_k\{S|y_k\} \! \prod\nolimits_{n=1}^N  
p^n_k( x_k^n | S,y_k  )   ,
\end{equation}
where $p^n_k( x_k^n | S, y_k)$ is the Kalman updated posterior Gaussian for target $n$ under assignment $S$, and
$\text{Pr}_k\{S | y_k\}$ is the posterior probability of assignment $S$; see Sect. \ref{resultsRecur}. 
To close the Bayesian recursion,
the multitarget posterior PDF is approximated as the product of
$N$ independent Gaussians in the same form as \eqref{eq:jpdaPrior}. In JPDA, this is done by replacing each posterior marginal, which is  a
Gaussian mixture,
with a single multivariate Gaussian with the same mean and covariance.


\section{The Ising Model for MTDA}\label{isingSec}

\subsection{Optimization Variables as Qubits}

Following Refs. \cite{mtda,fraunhofer2}, we may write each entry of the association matrix $S_{ij}$ of Eqn. (7) in Ref. \cite{aeroMtda} as the state of a two-level quantum system $\ket{\phi_{ij}} \in \{\ket{0},\ket{1} \}$. The presence or absence of an association is represented as one of the basis states of the two-dimensional complex Hilbert space given by
\begin{equation}
    \label{spinHalfBasis}
    \ket{0} = \begin{pmatrix}
           1 \\
           0 
         \end{pmatrix},\ 
     \ket{1} = \begin{pmatrix}
       0 \\
       1 
     \end{pmatrix},
\end{equation}
which are eigenvectors (with eigenvalues $1$  and $-1$, respectively) of the Pauli matrix $\sigma_3$, which in this basis is given by $ \sigma_3 = \begin{pmatrix}
           1 & 0\\
           0 & -1 
          \end{pmatrix}$.
The full state for the system in Eqn. (7) in Ref. \cite{aeroMtda} is the Kronecker product
\begin{equation}
    \label{prodState}
    \ket{\Phi_j} = \ket{\phi_{00}} \otimes \cdots \otimes \ket{\phi_{MN}},
\end{equation}
where $j$ is the integer corresponding to the binary string of association matrix entries:
\begin{equation}
    \label{binaryString}
    j \longleftrightarrow (S_{00}, S_{01}, \ldots, S_{MN}).
\end{equation}
In this way, all possible (feasible and infeasible) associations can be written as $D = 2^{(M+1)(N+1)}$ dimension quantum states formed from Kronecker products of $\sigma_3$
eigenstates.

\subsection{Optimization Problems as Ising Models}

Determining high-likelihood configurations of association variables is a manifestly classical problem. Valid configurations are given only by tensor products of Eqn. (\ref{spinHalfBasis}) and hence the cost of a particular configuration of association variables can be written only in terms of powers of mutually commuting $\sigma_3$ operators.  We restrict our discussion to configuration costs that are at most quadratic in the association variables such that all possible costs are given by eigenvalues of the Hamiltonian 
\begin{equation}
    \label{isingHam}
    H_P = \sum\nolimits_{i,j = 1}^{N_s} Q_{ij}\sigma^i_{3}\sigma^j_{3} + \sum\nolimits_{i=1}^{N_s} q_i \sigma^i_{3}\,.
\end{equation}
In Eqn. (\ref{isingHam}), $Q_{ij}$ captures interaction costs while $q_i$ captures ``biases" on all $N_s$ individual variables.  The operator $\sigma^i_{3}$ is  
\begin{equation}
\label{sigz_kron}
     \sigma^i_{3} = \underbrace{\sigma_0 \otimes\cdots\otimes \sigma_0}_{i-1\,\textrm{ terms}}   \otimes\, \sigma_3 \otimes \underbrace{\sigma_0 \otimes\cdots\otimes \sigma_0}_{N_s-i\,\textrm{ terms}}\, ,
\end{equation}
where $\sigma_0$ is the $2\times2$ identity matrix.

The Hamiltonian in Eqn. (\ref{isingHam}) is that of an Ising model, and it can be thought of as a finite-dimensional matrix whose entries are energies in the quantum mechanical sense that correspond exactly with the cost of association in the optimization problem of interest.  However, finding the equivalent optimal configurations exhaustively would require a number of trials that scales exponentially as $2^{N_s}$.  In the next section, we describe a quantum method to efficiently solve for low-cost configurations using a quantum annealer.  
In Sections \ref{resultsSingle} and \ref{resultsRecur}
we describe precise forms of $Q_{ij}$ and $q_i$ for two related problems of interest.

\section{Diabatic Quantum Annealing}\label{annealSec}

\subsection{The Instantaneous Hamiltonian}

In quantum annealing, the system at the instantaneous time $t$ is governed by the Hamiltonian
\begin{equation}
    \label{adiabham}
    H(t) = A(t) H_B + B(t) H_P\,,\quad t_0\le t\le t_f\,,
\end{equation}
where $A(t)$ and $B(t)$ are real-valued time-dependent coefficients.  The functional form of these coefficients comprise an \textit{annealing schedule}.  In the simplest case, which we refer to as \textit{forward annealing}, $A(t)$ monotonically decreases to 0 and $B(t)$ monotonically increases to 1.  

Throughout this work, we will only consider $H_B$ of the form 
\begin{equation}
    \label{dwaveinit}
    H_B = -\sum\nolimits_{i=1}^{N_s} \sigma^i_{1}\,,
\end{equation}
where, in analogy with Eqn. (\ref{sigz_kron}), 
\begin{equation}
\label{sigx_kron}
     \sigma^i_{1} = \underbrace{\sigma_0 \otimes\cdots\otimes \sigma_0}_{i-1\,\textrm{ terms}}   \otimes\, \sigma_1 \otimes \underbrace{\sigma_0 \otimes\cdots\otimes \sigma_0}_{N_s-i\,\textrm{ terms}}\, .
\end{equation}
Here $\sigma_1$ is the Pauli matrix  $ \sigma_1 = \begin{pmatrix}
           0 & 1\\
           1 & 0 
    \end{pmatrix}$. 
Since $\sigma_1$ and $\sigma_3$ do not commute, it is clear that, in general, $H_B$ and $H_P$ do not commute.  Therefore, $H(t)$ is not diagonal in the computational basis given by Eqn. (\ref{spinHalfBasis}), except when    ${A(t)=0}$.  The off-diagonal elements induce transitions between configurations in the computational basis.

\subsection{Time Evolution in Quantum Annealing}
 
At $t = 0$, the system is prepared in an initial state that we denote $\ket{\psi(0)}$.  During the course of the quantum annealing process, physical parameters of the device are modified following an annealing schedule $\{A(t), B(t)\}$ such that at any point in time $t$ the system evolves according to 
\begin{equation}
    \label{genTimeEv}
    \ket{\psi(t)} = U(t,0)\ket{\psi(0)}.
\end{equation}
The solution of this system of time-dependent ordinary differential equations is the quantum time-evolution operator $U(t,0)$.  It is written 
\begin{equation}
    \label{propagator}
	U(t,0) = \texp \textstyle{\left( -i \int_{0}^{t} dt' \, H(t')  \right)},
\end{equation}
where $H(t')$ is the quantum Hamiltonian and $\texp(\cdot)$ denotes the time-ordered exponential function which, expanded as a power series, is 
\begin{multline}
    \label{texp}
	\texp \textstyle{\left( -i \int_{0}^{t} dt' \, H(t')  \right)}\\
	= \sum_{n=0}^{\infty} (-i)^n\int_{0}^{t}dt_{n}'\cdots\int_{0}^{t_{2}'}dt_{1}'\, H(t_{n}')\cdots H(t_{1}'),
\end{multline}
where  $t > t_n > t_{n-1} > ... > t_1$.  Only in the specific case when the Hamiltonian commutes at different times, i.e.,  
\begin{equation}
    \big[H(t_{1}), H(t_{2})\big] = 0,\,\, \forall\, t_1, t_2,
\end{equation}
does the time-ordered exponential reduce to the more familiar form $U(t,0) = \exp\big(-i\int_{0}^{t}dt'  H(t')\big)$.

\subsection{Diabatic Transitions}

We write the Hamiltonian in Eqn. (\ref{adiabham}) as ${\mathcal{H}}(s) \equiv H(st_f)$, where $t_f$ is the final evolution time and $s \equiv t/t_f \in [0,1]$ is dimensionless. The paradigm of \textit{adiabatic} quantum annealing is to begin in a known and simple-to-prepare state $\ket{\psi_{0}(0)}$, often taken to be the ground state of $H_B$, and evolve the system \textit{slowly}. The
evolution of the system with Hamiltonian (\ref{adiabham}) is adiabatic if  the
``energy gap'' $|E_n(s) - E_m(s)|$ between any two instantaneous eigenstates of ${\mathcal{H}}(s)$ (with labels $n$ and $m$) is sufficiently large compared to the rate of evolution  \cite{amin_adiab}:  
\begin{equation}
    \label{adiabThm}
    \frac{1}{t_f} \max_{s\in [0,1]} \frac{|\bra{n(s)} \partial_s {\mathcal{H}}(s)  \ket{m(s)}|}{|E_n(s) - E_m(s)|} \ll 1, \forall m \neq n.
\end{equation}
Under this condition, the time evolution given by Eqn. (\ref{propagator}) simplifies such that the system remains in the ground state at all times and thus $\ket{\psi_{0}(t_f)}$ is the ground state of $H_P$.

For the purposes of tracking and multi-target data association, the lowest cost ``optimal" configuration (the ground state) may be a relatively low probability configuration due to a large  number of other configurations  that have  slightly higher energies. (This abundance of low-energy but non-ground state configurations is related to the fact that the difficulty of the MTDA problem increases rapidly as function of the number of targets and the false alarm density.) The Bayesian paradigm addresses this difficulty head-on by asserting, boldly, that the fundamental quantity of interest is the posterior distribution.  Point estimators are ancillary statistics derived from the posterior. 

Target state estimates used in this paper  are   expected values of the  posterior distributions.   
The expectations are estimated by summing over the feasible configurations, weighted by their likelihoods.  We use DQA to find the   ``low-energy" states to use in the Bayesian filter.  Consequently,  for our purposes, it is actually desirable to violate the condition in Eqn. (\ref{adiabThm}) in order to anneal to states which are higher in energy than the ground state.  

\subsection{Annealing to Excited States}

When a violation of adiabaticity occurs, quantum transitions are induced from the ground state into excited states.  In practice, since any annealing on a physical device must occur with a finite $t_f$, these transitions occur with relatively high probability.  This was noted in previous work on the application of quantum annealing to the MTDA problem \cite{aeroMtda}.  Too high an annealing rate, however, leads to a sequence of scatterings to high-energy states throughout the annealing process.   
Consequently, for sufficiently small anneal times $t_f$, 
excited states not relevant to the tracking and association process (due to their infeasibility) are populated with high probability.  Hence, we seek an intermediate range of values for $t_f$ such that only the ground state and low-energy feasible excited states are populated with  high probability.

 \begin{figure}[t]
    \centering
	\includegraphics
	[width=0.5\textwidth]
	{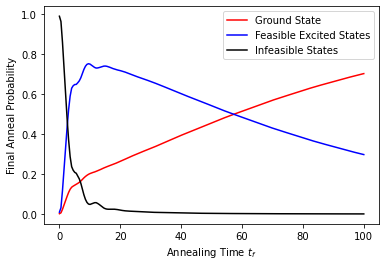}
	\caption {Final annealing probabilities for biased $k=3$ rooks problem obtained by classical numeric computation  QuTiP.  Units of anneal time are inverse energy.  A broad interval is observed for $t_f$ where the feasible excited states are annealed to.}
	\label{exact_diabatic_time_dependence}
\end{figure}

The eigenstates of the final Hamiltonian $H_P$ form a complete basis which we may use to expand any arbitrary state.  We may thus expand the final state after annealing 
\begin{equation}
    \ket{\psi(t)} = \sum\nolimits_{j = 0}^{D-1} c_{j}(t) \ket{\Phi_j},
\end{equation}
where $D = 2^{N_{s}}$ and $\ket{\Phi_j}$ for our problems of interest are given by Eqn. (\ref{prodState}).  The transition probability amplitudes from the initial state $\ket{\psi(0)}$ to one of the states in the computational product space $\ket{\Phi_j}$ (after an annealing time $t_f$) are given by
\begin{equation}
    \bra{\Phi_j}U(t_f,0)\ket{\psi(0)} = \braket{\Phi_j | \psi(t_f)}.
\end{equation}
From orthogonality of the eigenstates, the probability of annealing from $\ket{\psi(0)}$ to a state $\ket{\Phi_j}$ is given by 
\begin{equation}
    p_j = |\braket{\Phi_j | \psi(t_f)}|^2 = |c_{j}(t_f)|^2.
\end{equation}
A direct calculation of the transition amplitudes through solution of Eqns. (\ref{genTimeEv}-\ref{propagator}) depends on the precise form of the initial state $\ket{\psi(0)}$, the final problem Hamiltonian $H_P$, and the anneal schedule $\{A(t), B(t)\}$.  Obtaining a closed form solution is impossible in all but the simplest cases.  However, for small system sizes, we can numerically calculate the spectrum exactly by diagonalizing the Hamiltonian on a classical computer.  Using the exact eigenspectrum, we numerically calculate the time evolution of the dynamics of a simple system using the GKSL master equation \cite{intro2Lindblad,weinbergGKSL,gkslG,gkslL} in QuTiP \cite{qutip1,qutip2}.

In Fig. \ref{exact_diabatic_time_dependence} we show annealed occupation probabilities for the ground state, the feasible excited states, and the infeasible excited states as a function of the final anneal time $t_f$ for a biased $k=3$ rooks problem.  We take the parameters of the final problem Hamiltonian $H_P$ to be given by Eqns. (\ref{krooksrbaaa}) and (\ref{krookBiasLinearTerm}).  We take the bias term to be drawn from a normal distribution $q_b \sim {\mathcal{N}}(0,0.01)$.  We see that for very small anneal times, the final annealed states are dominated by higher-energy infeasible states (black curve).  Conversely, for very large values of $t_f$, the ground state probability is largest, as predicted by Eqn. (\ref{adiabThm}). For an intermediate range of $t_f$, we see that the final annealed states comprise both the ground state as well as feasible excited states while the probability of annealing to one of the infeasible states becomes small, despite their large number $(2^{k^2} - k!)$.

\begin{figure*}[!t]
    \centering
    \includegraphics[width=\textwidth]{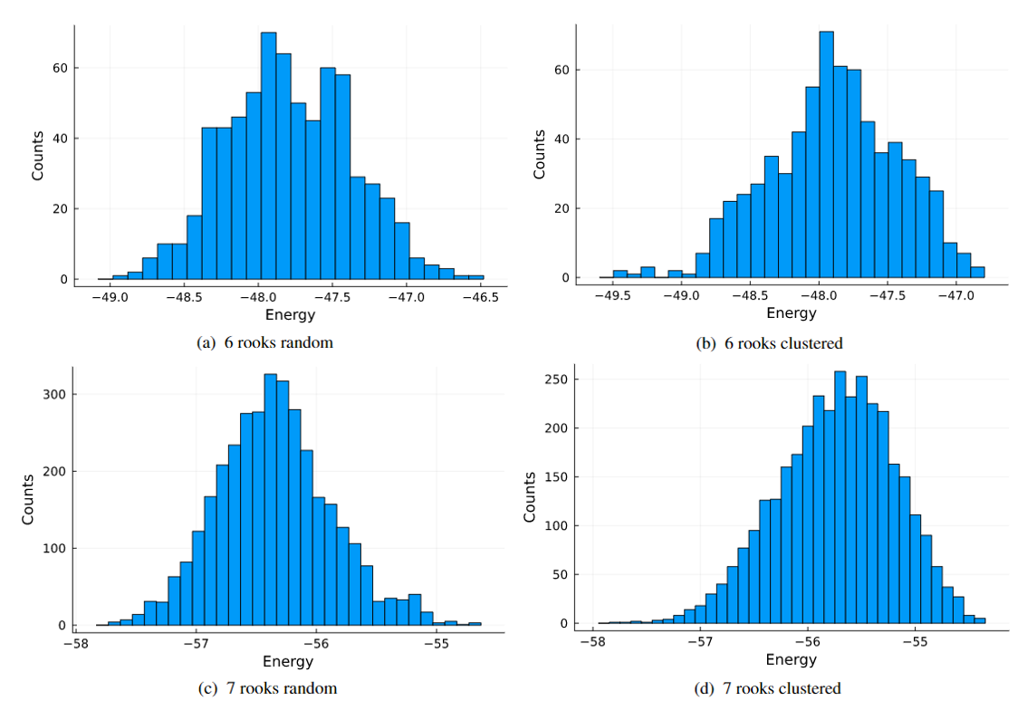}
    \caption {Unique feasible states found in $10^6$ shots on D-Wave 2000Q for different  $k$-rooks variations}
    \label{histograms}
\end{figure*}

\section{Diabatic Quantum Annealing for Feasible Configurations}
\label{resultsSingle}

In the previous section, by solving the GKSL master equation we showed that we theoretically expect diabatic quantum annealing to obtain feasible, low-energy states corresponding to low-cost combinatorial assignments in the $k$-rooks problem.  In this section, we validate this theoretical prediction using a D-Wave 2000Q QPU and study the how the fraction of feasible states that we obtain for a given number of quantum anneals depends on the system size.  We also show that the performance of diabatic quantum annealing depends strongly on the details of the anneal schedule itself.

\subsection{The Biased $k$-rooks Hamiltonian}

In the absence of false alarms and missed detections, the MTDA problem can be modeled as a $k$-rooks problem \cite{k-rooks}. In the $k$-rooks problem, a $k\times k$ chessboard is  populated with $k$ mutually non-threatening rooks.  We wish to represent the Ising form $H_P$ of the $k$-rooks problem in order to energetically impose the constraint of one rook per column and row.  Following the notation of \cite{k-rooks} and \cite{aeroMtda}, we define the following
\begin{align}
  {\mathbf{1}}(k) &\equiv k \times 1  
     \text{ column vector of ones} \nonumber  \\
          &=  \begin{pmatrix} 1 & 1 & 1 & \cdots & 1 \end{pmatrix}^T    \label{oneKcolVecrba}  \\
   I(k) &\equiv k \times k \text{ identity matrix}    \label{identityKrba} \\
   J(k) &\equiv {\mathbf{1}}(k) {\mathbf{1}}(k)^T - I(k)  \nonumber  \\ 
   &=  k \times k \text{ matrix of ones, with zero diagonal}   \nonumber  \\
  &=   \begin{pmatrix} 0 & 1 & 1 & \cdots & 1 \\ 
   1 & 0 & 1 & \cdots & 1 \\
      1 & 1 & 0 & \cdots & 1 \\ 
      \vdots & \vdots & \vdots & \ddots & \vdots \\ 
            1 & 1 & 1 & \cdots & 0 \\  
            \end{pmatrix}.   
\end{align}
The matrix of the quadratic term in the Ising Hamiltonian in Eqn. (\protect\ref{isingHam}) for the $k$-rooks problem is given by
\begin{equation}  \label{krooksrbaaa}
    \qqkrooks = \WrFromKrooks   +   \WcFromKrooks  \,,  
\end{equation}
and the linear term is given by
\begin{equation}     \label{krookLinearTerm}   
\qkrooks =  \thetaRKrooks + \thetaCKrooks  \, .   
\end{equation}
The $k^2 \times k^2$ matrix 
\begin{equation}     \label{qweqweA}
\WrFromKrooks \equiv I(k) \otimes J(k)
\end{equation} 
and the 
$k^2 \times 1$ column vector 
\begin{equation}  \label{qweqweB}
\thetaRKrooks \equiv (2k-4){\mathbf{1}}(k^2)
\end{equation} 
constrain the rows to have one rook each.
Similarly, 
\begin{equation}   \label{qweqweC}
\WcFromKrooks \equiv J(k) \otimes I(k)
\end{equation} 
and 
\begin{equation}   \label{qweqweD}
\thetaCKrooks \equiv (2k-4){\mathbf{1}}(k^2)
\end{equation} 
constrain the columns to have one rook each.  

The Ising Hamiltonian for the $k$-rooks problem described by Eqns. \eqref{krooksrbaaa}-\eqref{krookLinearTerm} has $k!$ degenerate ground states corresponding to the $k!$ different feasible configurations of rooks. 
Previous work \cite{aeroMtda}  demonstrated that these ground states can be obtained by a quantum annealer.  The high ground state degeneracy of the $k$-rooks Hamiltonian reflects the high degree of symmetry of the underlying problem.  

This degeneracy can be broken by the addition of a bias to the linear term.  We define \begin{equation}     \label{krookBiasLinearTerm}   
q_{kRB} =  \qkrooks + q_b,   
\end{equation}
where $ q_b$ is a $k^2 \times 1$ column vector whose elements correspond to costs for each space on the chess board (or equivalently to an on-site bias in the Ising Hamiltonian $H_P$).  We find that the addition of this bias term leads to a structured low-energy manifold of feasible states.

\subsection{Feasible configurations for the $k$-rooks problem}

\begin{figure}[!t]
    \centering
	\includegraphics[width=0.48\textwidth]{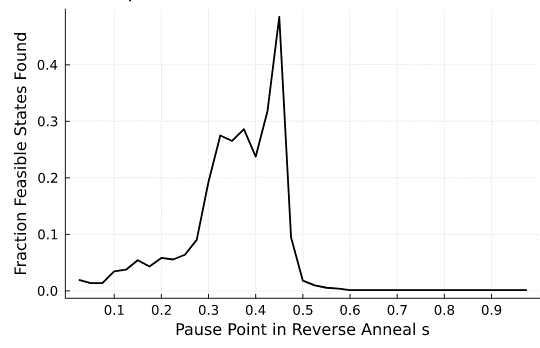}
	\caption {Dependence on anneal schedule of the number of unique feasible states found in $10^4$ shots for ${k = 6}$ rooks and random Gaussian bias.}
	\label{pausePoint}
\end{figure}

In this section, we obtain the low-energy costs and configurations of the $k$-rooks and MTDA systems using a D-Wave 2000Q QPU. For the $k$-rooks experiments we specify a chain strength of 8.  To  stimulate more rapid mixing to low-energy configurations, we use reverse annealing with an initial configuration of rooks along the diagonal and a pause-and-quench \cite{anneal_sched,rev_anneal} annealing schedule. Specifically, we anneal to the pause point at $s=0.45$ in 5 $\mu s$, pause at that point for 93 $ \mu s$, and then quench to the final Hamiltonian in 1 $\mu s$ (for a total anneal time of 99 $\mu s$). Fig. \ref{pausePoint} demonstrates the  potential utility of  reverse annealing and different pause points for a small number of shots ($10^4$).  See the last paragraph of this section for further discussion. 

\begin{figure}[!t]
    \centering
    \includegraphics[width=.48\textwidth]{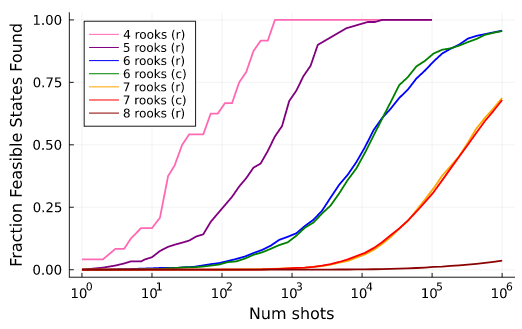}
    \caption {Unique feasible states accumulated over shots for different configurations of $k$-rooks (r = random bias, c = clustered)}
    \label{scaling}
\end{figure}

In Fig. \ref{histograms}, we show the the distribution of unique feasible states that D-Wave finds in $10^6$ shots for different variations of the $k$-rooks problem. 
These histograms show the density of annealed states.  As the number of shots is increased, the probability of obtaining all feasible states empirically is found to approach unity (see Fig. \ref{scaling}). The counts in Fig. \ref{histograms} correspond to the total unique feasible states found.  Generating a good solution to the MTDA problem relies on finding as many of the feasible states as possible, but the frequency with which each of those states occur in the returned shots is not relevant.

In Figs. \ref{histograms} and \ref{scaling}, we call certain $k$-rooks variations ``random'' (``r'') or ``clustered'' (``c''). This refers to the on-site bias ($q_b$ in Eqn. \ref{krookBiasLinearTerm}) applied to each square of the $k$x$k$ board. In the random variations, the bias is drawn from a Gaussian distribution $q_b \sim {\mathcal{N}}(0,0.1)$ homogeneously across the entire board. 
For the clustered variations, the signs of the randomly drawn biases are adjusted so that certain blocks along the diagonal have considerably smaller bias (i.e., contribute lower energy) than squares outside those blocks. Since the $k$-rooks system is equivalent to the MTDA problem with no misses or false alarms, this biased variation models a system where there are several clusters of targets and measurements which are harder to disambiguate than those from cluster to cluster.   Feasible states that respect the block structure have lower energy than others, as reflected in the longer left-tail in the (b) and (d) histograms of Fig.  \ref{histograms}. For the 6-rooks case the block sizes are $[1,2,3]$ and for the 7-rooks case they are $[2,2,3]$.

Fig. \ref{scaling} shows, for the $k$-rooks problem, the fraction of feasible states found by quantum annealing as a function of the number of shots $n_s$ for several different values of $k$ and for different bias variations.  Although the proportion of total states that are feasible ($k!/2^{k^2}$) decreases rapidly as $k$ increases, we find a remarkably large fraction of the possible feasible states: for $6$-rooks, feasible states represent about $10^{-8}$ of the total states; for $7$-rooks, the fraction drops to \num{9e-12}.  For both $k = 6$ and $k = 7$, the number of shots required for each bias variation (random vs. clustered) is very similar. Fig. \ref{scaling} demonstrates for the $k$-rooks problem that DQA will find the feasible configurations even when they are like   needles ``lost'' in a haystack of infeasible configurations.

As noted in the previous section, anneal parameter selection plays a critical role in obtaining as large a fraction of feasible states as possible.  Using a reverse anneal pause-and-quench schedule, we found that the fraction of feasible states found depended strongly on the pause point (Fig. \ref{pausePoint}).  Fig. \ref{pausePoint} shows a slow rise as $s$ increases to a sharp peak at $s=.45$ and an immediate fall-off afterwards in the fraction of feasible states found. In Fig. \ref{pausePoint}, we run anneals of 10,000 shots on a D-Wave 2000Q for a $6$-rooks problem with random on-site bias, for every pause point in a grid of resolution 0.025.

\section{Multistep Bayesian Recursion with Quantum Annealing}
\label{resultsRecur}

In this section 
the DQA algorithm discussed in Sec. \ref{annealSec} is used recursively to find  feasible assignments.  The DQA generated assignments are passed to a classical computer to find the track estimates via the JPDA recursion given in Sec. \ref{dataassoc}.  The recursion is closed by passing the estimated tracks back to the DQA algorithm to find feasible assignments for the next scan. To our knowledge, this is the first demonstration of a  Bayesian hybrid quantum-classical multiple target tracking filter.  

\subsection{The MTDA Hamiltonian}

The Ising formulation of the MTDA problem comprises two different components in the problem Hamiltonian $H_P$. The first is a quadratic term that constrains the association of at most one target to each detection and at most one detection to each target. The second is a linear term that accounts for the cost of target-detection misassignment based on the negative log-likelihood.  The linear term also contains the linear part of the association constraints, similar to Eqn. \eqref{krookLinearTerm}.
Let $\gamma = \text{vec}(\costQ)$ be as defined by Eqn. (9) from our previous work \cite{aeroMtda}, 
and let $c > 0$ be given.
Then the quadratic term that constrains the association of at most one target to each detection and at most one detection to each target
leads to a modified form of the $k$-rooks cost matrix
\begin{equation} \label{mtda_cost}
 \qqmtda = c \, (\WrFromMTDA + \WcFromMTDA), 
\end{equation}
and the second term gives
\begin{equation}     \label{mtda_cost_linear}
    \qmtda = c \, (\thetaRMTDA + \thetaCMTDA) + \gamma  \, ,   
\end{equation}
where $\WrFromMTDA$, $\WcFromMTDA$, $\thetaRMTDA$, and $\thetaCMTDA$
(defined below) are similar 
to the corresponding $k$-rooks constraints 
but are modified to allow for missed detections and false alarms. 
The cost constraint coefficient $c$ 
sets the overall energy scale of the constraint terms;
higher values correspond to more stringent enforcement of the feasibility constraints. 

\begin{figure*}[!ht]
    \centering
    \includegraphics[width=\textwidth]{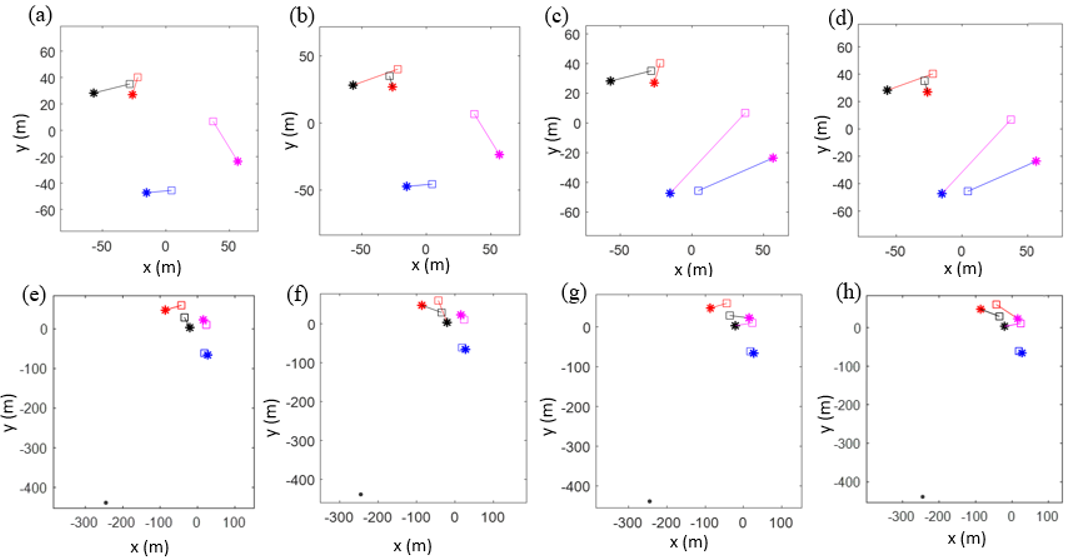}
    \caption{Assignments from Multi-Step MTDA performed using DQA. $\square$ = target, $\ast$ = measurement, $\bullet$ = clutter.  \textbf{(a)} Scan 1, Assignment 1, weight = 0.49596. \textbf{(b)} Scan 1, Assignment 2, weight = 0.42746. \textbf{(c)} Scan 1, Assignment 3, weight = 0.02806. \textbf{(d)} Scan 1, Assignment 4, weight = 0.02418. \textbf{(e)} Scan 2, Assignment 1, weight = 0.73170. \textbf{(f)} Scan 2, Assignment 2, weight = 0.12453. \textbf{(g)} Scan 2, Assignment 3, weight = 0.10950. \textbf{(h)} Scan 2, Assignment 4, weight = 0.02210.}
    \label{multistep}
\end{figure*}

With minor changes to Eqns. (\ref{oneKcolVecrba})  and (\ref{identityKrba}), 
we define the following matrices: 
\begin{align}
   {{\mathbf{1}}_0}(k) &\equiv k \times 1  
     \text{ col vector of ones with zero in first entry} \nonumber  \\
          &=  \begin{pmatrix} 0 & 1 & 1 & \cdots & 1 \end{pmatrix}^T    \label{oneKcolVecrbaZero}    \\         
    I_0(k) &\equiv k \times k \text{ identity matrix with zero in } (1,1) \text{ entry}  \nonumber  \\
  &=  \begin{pmatrix}   
  0 & 0 & 0 & \cdots & 0 \\ 
   0 & 1 & 0 & \cdots & 0 \\
      0 & 0 & 1 & \cdots & 0 \\ 
      \vdots & \vdots & \vdots & \vdots & \vdots \\ 
            0 & 0 & 0 & \cdots & 1 \\  
            \end{pmatrix}             \label{identityKrbaZero}
\end{align}

For $N \geq 1$ targets and $M \geq 0$ measurements,
\begin{align}
\WrFromMTDA &= I_0(M+1) \otimes J(N+1)   \nonumber  \\
\WcFromMTDA &= J(M+1) \otimes I_0(N+1)   \nonumber  \\
\thetaRMTDA &= \left(2N-2\right)\Big[{{\mathbf{1}}_0}(M+1) \otimes {\mathbf{1}}(N+1)\Big]    \nonumber \\
\thetaCMTDA &= \left(2M-2\right)\Big[{\mathbf{1}}(M+1) \otimes {{\mathbf{1}}_0}(N+1)\Big]     \,.    \label{mtdarbaaa}
\end{align}
We note that the MTDA constraint matrices 
in  Eqn. (\ref{mtdarbaaa}) provide the correct expression of the matrices given in \cite{mtda}. 

\subsection{Results for Hybrid Quantum/Classical DQA/JPDA}

We consider an $N = 4$ target tracking problem over multiple time scans.  We take the PPP clutter model described in Sec. \ref{dataassoc} with $\lambda = 1$ mean clutter measurements per scan.
The prior object state for each target $n$ is instantiated at reference 
time $t_1=1$ rather than $t_0=0$. The prior PDF is assumed to be normally distributed with mean $\hat{x}^n_{{{1 \givennn 1}}}$ equal 
to the corresponding ground truth at time $t_1$ and
diagonal covariance $P_{1 \givennn 1}^n = 
\text{diag}(25^2,5^2,25^2,5^2)$; 
see Eqn. (\ref{eq:jpdaPrior}).
At each time scan, a cost matrix $\Gamma$  is derived from the posterior assignment $\text{Pr}_k\{S | y_k\}$ in Eqn. (\ref{eq:jpdaExactPost}) and converted into a linear bias $\gamma$ in Eqn. (\ref{mtda_cost_linear}).  The related MTDA model (with $c = 2$) is then solved on a D-Wave 2000Q QPU using $n_s = 10^4$ shots.  All feasible configurations (those satisfying Drummond’s ``at most one measurement per target per scan rule”) that are returned by the DQA are then passed back to the classical JPDA tracker.  They are then assigned weights according to $\text{Pr}_k\{S | y_k\}$ and used to generate the prior PDF 
for the next scan. 

In {Fig. \ref{multistep}} we show the measurements and ground truth locations for the top four assignments (rank-ordered by their posterior weight) for two consecutive scans.  Square icons show the ground truth locations, stars show the measurement locations for targets, and filled circles correspond to the locations of clutter measurements.  Colored lines connecting ground truth and targets show the assignments returned by the DQA algorithm.
The best scoring assignment for both scans assigns all of the targets and measurements properly.

\begin{figure}[!t]
    \centering
	\includegraphics[width=0.48\textwidth]{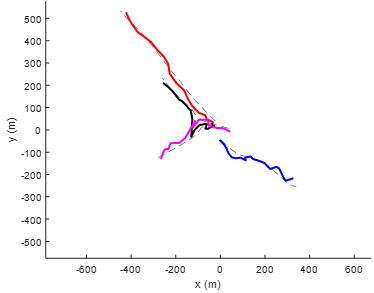}
	\caption {Hybrid DQA/classical track estimates for four targets by over 19 time steps. Target state  estimates are the expected values obtained  by summing  over the  DQA generated assignments.  Targets are initialized near the origin and move outward. $Pd=0.9$. False alarms are uniformly distributed and Poisson ($\lambda=1$). (dashed lines = ground truth)}
	\label{19scan}
\end{figure}

In Fig. \ref{19scan}, we show tracks of all 4 targets over 19 consecutive time scans. Targets begin clustered tightly nearly the origin and move outward (ground truth shown as dashed lines).  It is difficult to see in the figure, but the black and magenta targets cross roughly a quarter of the way through the simulation, making this a difficult tracking problem.  Given the measurement errors, the  JPDA tracker is not able to disambiguate these targets and assigns improper custody by the end of the scenario.  The DQA generated assignments correctly reflect this reality in the data.  Tracks on the blue and red targets are assigned properly.

\section{Concluding Remarks}
\label{ConcludingRemarks}

In this paper we have demonstrated that diabatic quantum annealing can be used to efficiently explore the space of low-energy states for a wide range of Ising model Hamiltonians that are relevant for multi-target data association problems.  These low-energy states correspond to nearly-optimal feasible assignments, and thus they can be summed over to compute Bayesian mean state estimators.  Moreover, as seen in Fig. \ref{19scan}, this can yield accurate results for a non-trivial tracking problem, thereby concretely establishing the utility of our novel hybrid quantum/classical approach to a wide class of tracking problems.  

One crucial point worth emphasizing is that the feasible low-energy states comprise only a tiny fraction of the full state space in the models explored in this work.  In particular, for the $k$-rooks problem with $k \geq 6$, fewer than one in every $10^8$ states represent feasible data associations, which makes classical approaches for finding such associations prohibitively expensive.  Our DQA-based approach, however, yields almost all of these feasible associations for the $k = 6$ and $k = 7$ problems.  Our approach to DQA-based feasible state identification will naturally extend to other models relevant for tracking problems, and we look forward to exploring these possibilities  in upcoming work.

The efficiency of using DQA to find feasible states hinges on the selection of annealing schedules and annealing times that break adiabaticity ``softly'' enough to avoid annealing to high-energy states.  As such, it would be worthwhile to develop a  systematic and rigorous approach for determining how best to select the annealing parameters for any given MTDA problem.  This can  be done  via a path integral Monte Carlo approach, wherein the full quantum time evolution of the Ising model is simulated by doing classical annealing on a large number of ``replica" copies of the original model \cite{quantum_monte_carlo}.  We leave such an analysis to future work.

\acknowledgments
This work was supported by Metron under corporate IRAD funding  as part of a larger Quantum Computing initiative.  We thank Martin Suchara, Sebastian Hassinger, and Mike Ashford at AWS for a computational time grant for this work.

\bibliographystyle{IEEEtran}

\thebiography
\begin{biographywithpic}
{Timothy M. McCormick}{mccormickt.png}
received his B.S. in physics from University of Delaware and his Ph.D. in theoretical physics from the Ohio State University, where his research focused on thermoelectric transport and electronic structure of topological semimetals. He is a research scientist at Metron Inc., where his work lies at the intersection of physics, statistics, and computation.  His current interests include underwater acoustics, compressed sensing, multi-agent reinforcement learning, and quantum computation.
\end{biographywithpic} 

\begin{biographywithpic}
{Zipporah Klain}{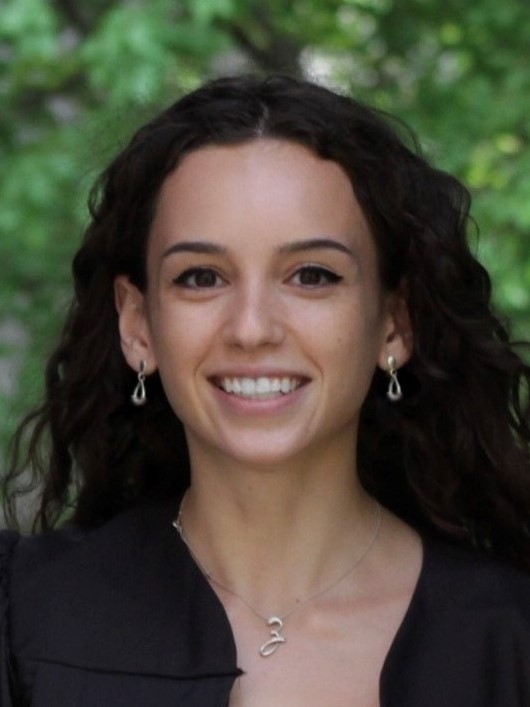}
received a B.S. in Computer Science and a B.A. in Sociology from the University of Chicago in 2021. While there, her research spanned several domains, including effective quantum-computing education for all levels of students with the EPiQC and CANON labs. She is currently a software engineer at Metron, Inc., working on research and production projects focused on tracking, automated mission planning and optimization, and applications of quantum computing.
\end{biographywithpic}

\begin{biographywithpic}
{Ian Herbert}{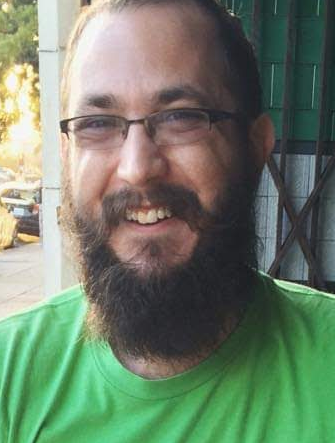}
received a B.S. in Mathematics and a B.A. in Russian from Tulane University, and a Ph.D. in Logic and the Methodology of Science from UC Berkeley. He is a research scientist in the Advanced Data Analytics division at Metron, Inc., where his work focuses on AI/ML, particularly adversarial-input attacks to ML systems and defenses against them. His current interests include Gaussian Processes, Graph Convolutional Networks, and anomaly detection.
\end{biographywithpic}

\begin{biographywithpic}
{Anthony M. Charles}{anthonypic.jpg}
received a B.S. in physics and a B.A. in mathematics from the University of Virginia and a Ph.D. in theoretical physics from the University of Michigan, where his research focused on quantum aspects of black holes in string theory.  He is currently a software engineer at Metron, Inc., where he works on research and development across a wide range of topics in physics and mathematics, including particle filtering and sequential Monte Carlo methods, optimal control theory, and quantum computing.
\end{biographywithpic}

\begin{biographywithpic}
{Bryan R. Osborn}{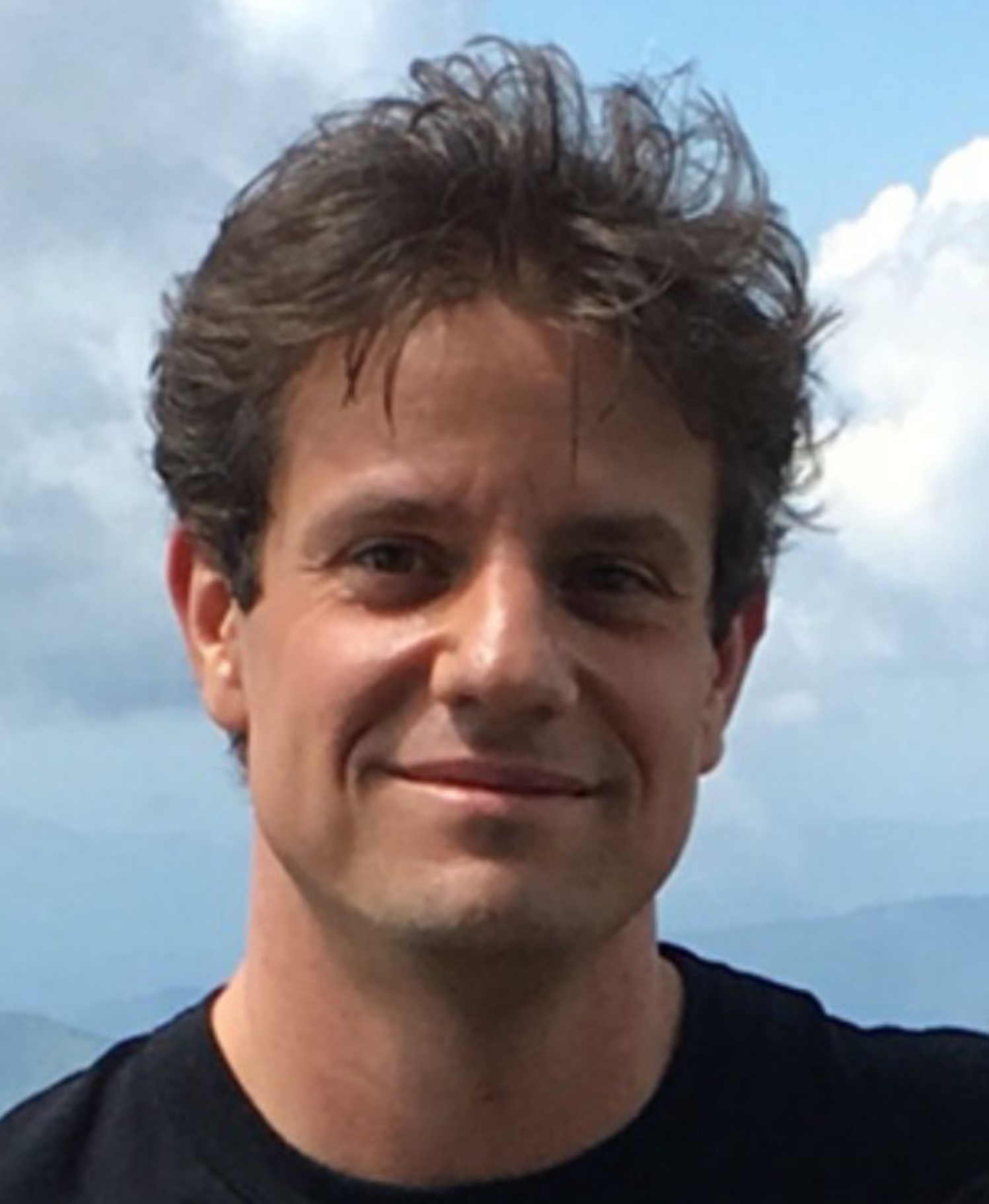}
received  B.S. degrees in Physics and Mathematics in 2001 and an M.S. in Applied Mathematics and Scientific Computation in 2004 from the University of Maryland, College Park.  He is currently a Senior Research Scientist at Metron, Inc. where he leads development and application of tracking algorithms in a variety of contexts.  His interests include distributed sensing systems, high-performance computing, and interactive visualization.
\end{biographywithpic}

\begin{biographywithpic}
 {R. Blair Angle}{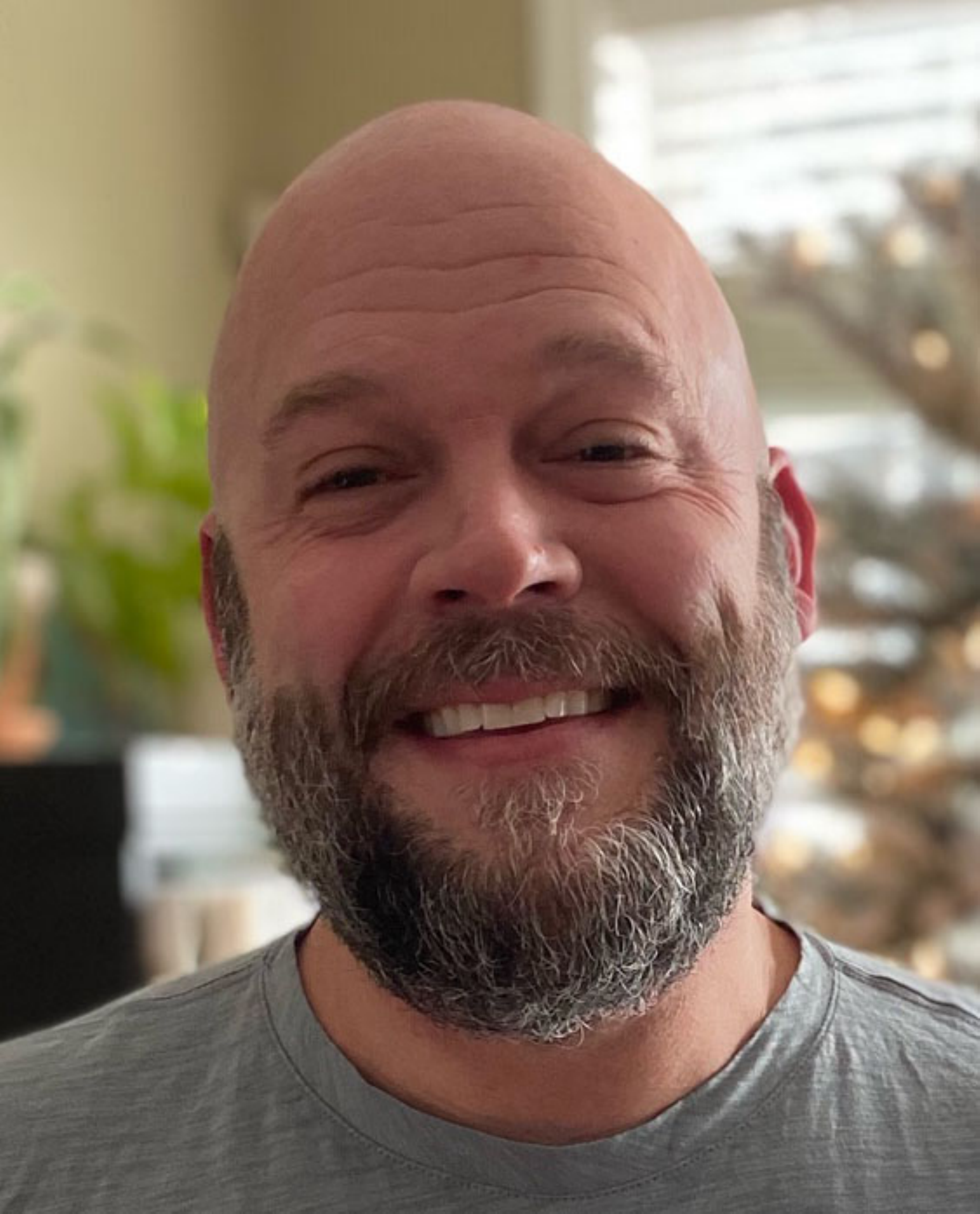}   
received  his Ph.D. in mathematics from the University of California, San Diego,
and is currently a senior research scientist at Metron, Inc.
His current work involves multi-target tracking, data fusion, analytic combinatorics,
and their interplay.
He recently co-authored the book Analytic Combinatorics for Multiple Object Tracking, Springer, 2021.
\end{biographywithpic}

\begin{biographywithpic}
{Roy L. Streit}{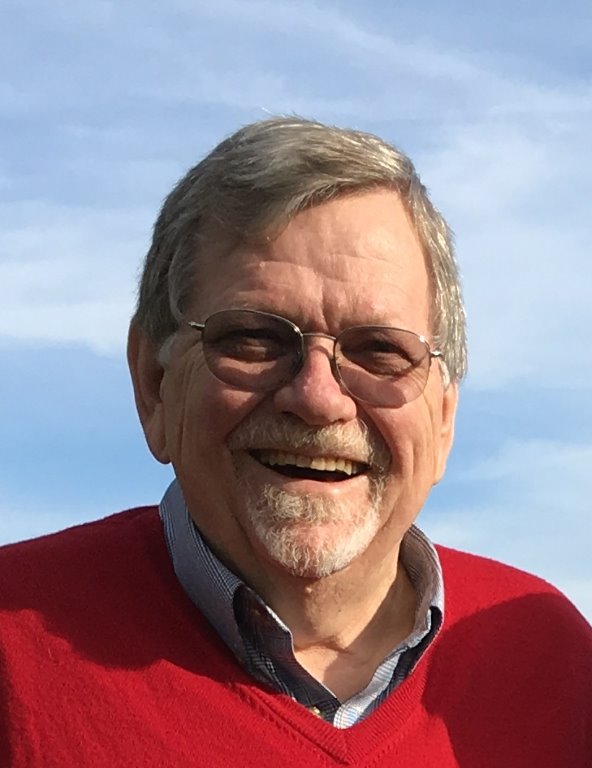}
joined Metron in 2005.  His interests include multi-target tracking, multi-sensor data fusion, signal processing, medical imaging, and quantum computing.  His recent work involves applications of analytic combinatorics to multi-target tracking, natural language processing, and subgraph matching in high level fusion.  He is a Life Fellow of the IEEE.  He co-authored Analytic Combinatorics for Multiple Object Tracking, Springer, 2021, and Bayesian Multiple Target Tracking, Second Edition, Artech, 2014.  He is also the author of Poisson Point Processes: Imaging, Tracking, and Sensing, Springer, 2010.  Before 2005, he was in Senior Executive Service at Naval Undersea Warfare Center, Newport, RI.
\end{biographywithpic}

\end{document}